\documentclass[prb,twocolumn,showpacs]{revtex4}
\usepackage{graphicx,hyperref}

\begin{document}
\title{Dissipation due to tunneling two-level systems\\ in gold nanomechanical resonators}
\author{A. Venkatesan}
\email{ananth.venkatesan@nottingham.ac.uk }
\author{K. J. Lulla}
\author{M. J. Patton}
\author{A. D. Armour}
\author{C. J. Mellor}
\author{J. R. Owers-Bradley}
\affiliation{School of Physics and Astronomy, University of Nottingham, Nottingham NG7 2RD, United Kingdom.}

\begin{abstract}
We present measurements of the dissipation and frequency shift in nanomechanical gold resonators at temperatures down to 10 mK.
The resonators were fabricated as doubly-clamped beams above a GaAs substrate and actuated magnetomotively. Measurements on beams with frequencies 7.95 MHz and 3.87 MHz revealed that from 30 mK to 500 mK the dissipation increases with temperature as $T^{0.5}$, with saturation occurring at higher temperatures. The relative frequency shift of the resonators increases logarithmically with temperature up to at least 400 mK. Similarities with the behavior of bulk amorphous solids suggest that the dissipation in our resonators is dominated by two-level systems.
\end{abstract}
\maketitle

Nanomechanical resonators with frequencies ranging from a few MHz up to 1 GHz have important potential applications as ultra-sensitive force sensors,\cite{ekinci} and are also ideal systems in which to study mesoscopic physical phenomena involving mechanical degrees of freedom. Coupling nanomechanical resonators electrostatically to mesoscopic circuit elements\cite{naik,lehnert} allows displacement detection with almost quantum-limited sensitivity to be achieved. Furthermore, when cooled to sufficiently low temperatures, nanomechanical resonators are expected to display quantum coherence.\cite{miles} A detailed understanding of the processes responsible for dissipation in nanomechanical resonators is crucial if the full potential of such devices is to be realized in device applications. Moreover, an understanding of how nanomechanical resonators dissipate energy at low temperatures is an important prerequisite for developing appropriate models of environmental decoherence in such systems.\cite{max,blencowe}

Despite significant experimental\cite{greywall,ekinci,mohanty,diamond} and theoretical\cite{ted,lc,iwr,ahn,seonaz,blencowe} efforts a full picture of the mechanisms that control the $Q$-factor of nanomechanical resonators has yet to be established, especially at low temperatures. Contributions to the mechanical dissipation will always come from clamping losses via the supports\cite{iwr,lc} and thermoelastic damping,\cite{ted} but for beam and cantilever resonators with large aspect ratios and at low temperatures, neither is expected to be the dominant factor. Rather, defects within the crystalline structure of the resonator, or on its surface, are expected to play an important role.\cite{ekinci} The low temperature dissipation in larger acoustic oscillators, especially those made from amorphous solids, has been studied extensively and is widely interpreted in terms of defects modeled as tunneling two-level systems.\cite{esqui} An important prediction of the standard tunneling model for amorphous solids is a characteristic power law behavior for the dissipation at very low temperatures, with a cubic dependence expected for dielectrics and a linear power for metals.\cite{esqui} In contrast, only a handful of studies of dissipation in nanomechanical resonators below 1 K have been conducted.\cite{greywall,mohanty} However, recent low-temperature measurements\cite{mohanty} on beam resonators with frequencies in the range 5--45 MHz fabricated from Si, GaAs and diamond have suggested a $T^{1/3}$ power law dependence for the dissipation.

Here we present the results of a study of dissipation in polycrystalline gold nanomechanical resonators cooled down to 10 mK. Metal resonators are being used increasingly in low temperature experiments since the fabrication is simplified when an underlying dielectric layer is not required\cite{li,lehnert,cooling} and the $Q$-factor of these resonators has been found to be relatively large, although the temperature dependence of the dissipation has not been studied systematically. As metal resonators are readily fabricated and actuated as monolithic structures, the interpretation of dissipation measurements is simpler than for dielectric ones where metallization is typically required to facilitate actuation.\cite{greywall,mohanty,diamond} A short account of some of our results for one of the gold resonators has been given elsewhere.\cite{venkatesan}

We fabricated nanomechanical resonators consisting of doubly-clamped gold beams using a combination of e-beam lithography and wet etching.\cite{venkatesan} Polycrystalline gold was deposited by thermal evaporation onto a $3$ nm thick adhesion layer of titanium and the beam released by wet etching the underlying GaAs substrate.  Our results were obtained using two beams: one with length $l=7.5$ $\mu$m, width $w=300$ nm, thickness $t=80$ nm and a second sample with $l=10.5$ $\mu$m, $w=250$ nm, $t=65$ nm. A SEM micrograph of the longer gold beam resonator is shown in Fig.\ 1a with a schematic diagram of the measurement set-up. Using the SEM we estimate that the grain size of the polycrystalline gold is 20--40 nm.

The $Q$-factor and frequency of the fundamental out of plane flexural modes of the beams were measured using  magnetomotive methods.\cite{cleland2} The spectral response of the beam was obtained using a continuous wave (CW) technique by applying a magnetic field (parallel to the plane of the wafer) and  passing a radio-frequency alternating current through the beam. The response to the resulting Lorentz force was detected by measuring the induced emf (using a heterodyne detection scheme) as the frequency of the current was tuned through the mechanical resonance. The frequency and $Q$-factor were then extracted by fitting the spectrum to a standard Lorentzian, after a linear background was subtracted. Excitation powers of -100dBm or less were used and care was taken to ensure that the resonator remained in the linear response regime throughout.

We also made measurements using a ring-down approach\cite{diamond} in which the resonator was excited by a 200 $\mu$s radio-frequency pulses {(spaced 1 s apart)} with a frequency close to the mechanical resonance and then the response measured in the time domain. Examples of the ring-down signal are shown in Fig.\ 1b; the $Q$-factor was extracted by fitting an exponential to the envelope of the resulting oscillations. Obtaining accurate estimates of the resonator frequency from the ring-down method was more problematic because of transient effects which mean that the oscillations come closest to the natural frequency at later times when the signal itself is rather small.

\begin{figure}[t]

\centering

\scalebox{0.7}{\includegraphics{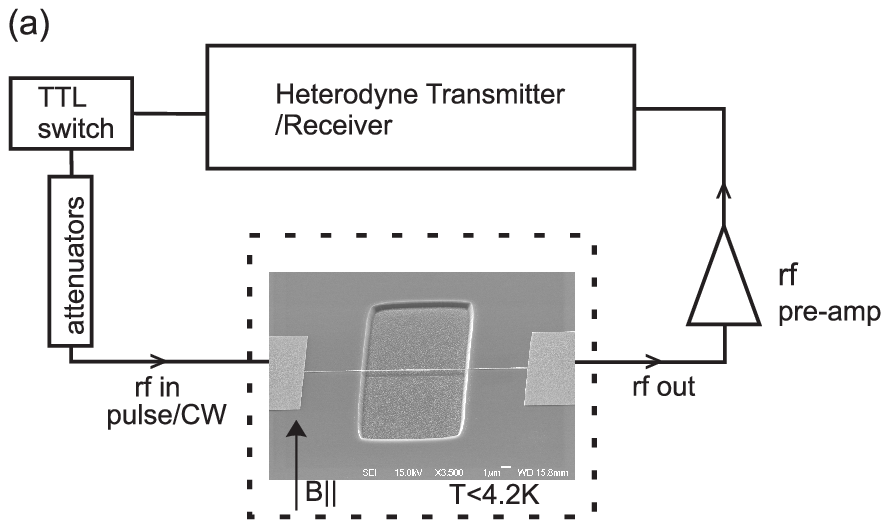}}
\scalebox{0.7}{\includegraphics{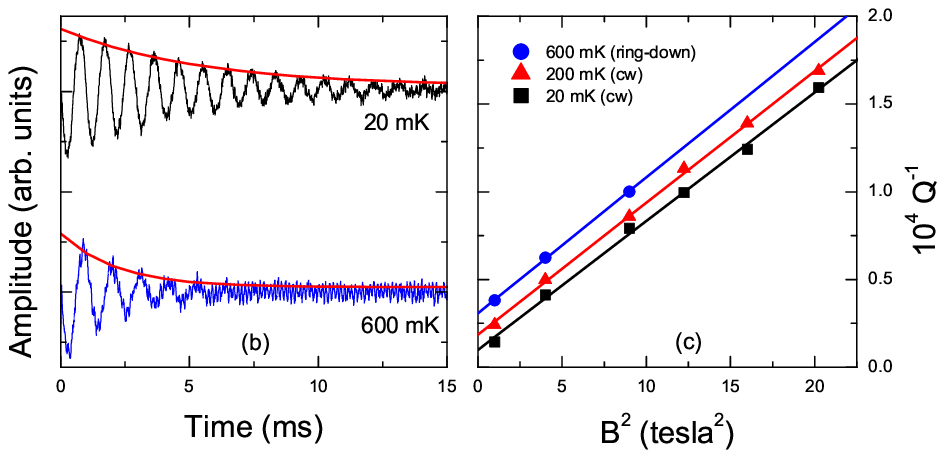}}

\caption{\footnotesize(a) Detection set-up used to perform the magnetomotive measurements. A SEM micrograph of one of the samples is also shown. (b) Typical  ring-down data for the $3.87$ MHz device taken at 1 T, the smooth lines are exponential fits to the envelope (the data sets are displaced for clarity). (c) $Q^{-1}$  versus $B^2$ for the $3.87$ MHz device at three different temperatures; the lines are linear fits to the data. }\label{setup}
\end{figure}

The resonant frequency of the beams at low temperatures is  affected significantly by differential thermal contraction between the GaAs substrate and the metal.\cite{li,lehnert} To a good approximation a doubly-clamped beam of length $l$ under tension $T_0$ has a fundamental frequency given by\cite{roukes,li}
\begin{equation}
f_0 = \frac{2\pi}{l^2}\sqrt{\frac{Et^2}{36\rho}+\frac{l^2T_0}{12\pi^2 \rho A}}           \label{tension}
\end{equation}
where $A=wt$ is the cross-sectional area, $E$ the Young's modulus and $\rho$ the density. The tension arising from differential thermal contraction  between GaAs and Au when the beam is cooled can be estimated using\cite{li} $T_0=EA[(\frac {\Delta l}{l})_{{\rm Au}}-(\frac{\Delta l}{l})_{{\rm GaAs}}]$, where $(\Delta l/l)_{{\rm Au}}$ is the fractional change in length of the gold on cooling.
Using Eq.\ \ref{tension} and simple estimates of the differential thermal contraction obtained from the literature\cite{li,augaas} we obtain estimates for the frequencies of 7.9 MHz and 5.4 MHz for the beams with lengths $7.5$ $\mu$m and $10.5$ $\mu$m respectively. The measured frequency for the shorter beam was $7.95$ MHz, but for the longer beam it was $3.87$ MHz, somewhat lower than our estimate.  This deviation is most probably due to initial compressive strain which we found (based on our experience of fabricating a number of resonators using the same process) to be generated in varying degree by the fabrication process.

Use of the magnetomotive transduction technique results in a measured $Q$-factor which is loaded by the external circuitry and the electrical resistance of the sample itself. However, the unloaded (intrinsic) $Q$-factor, $Q_0$, can be extracted from the magnetic field dependence of the measured response, $Q$.\cite{cleland2,diamond}  In our system, the drive and detection circuits each have a $50$ $\Omega$ characteristic impedance along with some real resistance due to the cables and sample. Using an effective circuit model for our experiment in which the resonator is modeled as a parallel LCR circuit, we find that (assuming the external impedances presented to the sample are effectively real)
\begin{equation}
{Q}^{-1} ={Q_0}^{-1}\left( 1+ \frac{R_m \Re(Z_{ext})}{\vert Z_{ext}^{2}\vert}\right),   \label{q}
\end{equation}
where $Z_{ext}$ is the external impedance and  $R_m = \frac{\zeta B^2 l^2 Q_0}{2\pi f_0m}$ is the effective resistance of the resonator with $\zeta$ a mode constant of order unity and $m$ the resonator mass. We measured the dissipation as a function of $B^2$ at a range of different temperatures. As expected the behavior is found to be linear and the gradient allows $Q_0^{-1}$ to be obtained by extrapolation to zero field. The results for the $3.87$ MHz beam are shown in Fig.\ 1c and it is clear that the magnetic field can cause a substantial change in the dissipation. We found that the gradients showed only a very weak temperature dependence (which we ascribe to fractional changes in the impedances). Hence we used a simple average of the measured gradients to infer $Q_0$ at arbitrary temperatures. No corresponding quadratic field dependence was found for the resonant frequencies of the samples confirming that the impedances seen by the sample can be treated as effectively real.
\begin{figure}[t]

\scalebox{0.7}{\includegraphics{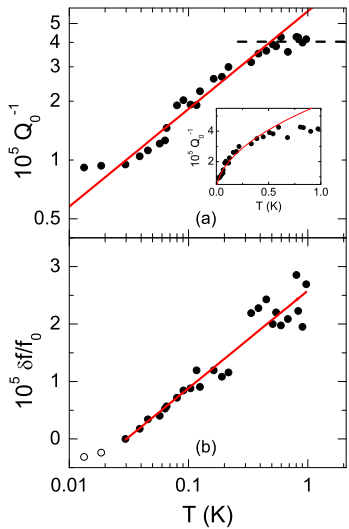}}
\caption{\footnotesize 7.95 MHz resonator: (a) $Q_0^{-1}$ as a function of temperature, on a log-log scale (main plot) and a linear scale (inset). The full line (curve in the inset) is a guide to the eye $\propto T^{0.5}$. The dashed line is the average of the five points with the highest temperatures. (b) Fractional frequency shift, $\delta f/f_0$, as a function of temperature. The line is a logarithmic fit to the data marked with full circles (i.e.\ excluding the two points below $30$ mK). Note that the data were collected in two runs and a frequency jump of 20.4 kHz that occurred during thermal cycling between the runs has been subtracted from the low temperature data.}

\end{figure}
\begin{figure}[t]

\scalebox{0.7}{\includegraphics{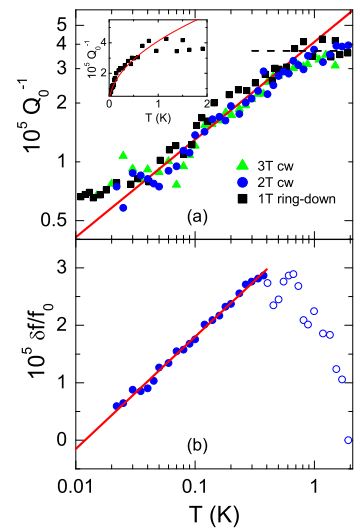}}
\caption{\footnotesize 3.87 MHz resonator: (a) $Q_0^{-1}$ as a function of temperature, on a log-log scale (main plot) and a linear scale (inset). The full line (curve in inset) is a guide to the eye $\propto T^{0.5}$. The dashed line is the average of the five points with the highest temperatures.  Only the 1 T ring-down data are shown in the inset for clarity. (b) Fractional  frequency shift, $\delta f/f_0$, as a function of temperature (CW data at 2 T). The line is a logarithmic fit to the sub-set of data points marked with full circles.}
\end{figure}

 We carried out a series of measurements of the $Q$-factor and frequency of the 7.95 MHz and 3.87 MHz resonators from about 1 K down to 10 mK. The data for the 7.95 MHz device were obtained from CW measurements at a field of 3 T, but for the 3.87 MHz sample CW measurements were made at 2 T and 3 T, as well as ring-down measurements at 1 T. The ring-down measurements at 1 T provided the most accurate results (as can be seen by comparing the scatter in the different data sets in Fig.\ 3a) since less extrapolation was required at this field strength to obtain $Q_0^{-1}$.

The measured temperature dependence of the intrinsic dissipation, $Q_0^{-1}$, for the resonators is shown in Figs.\ 2a and 3a respectively. The data from the samples follow a common pattern and three different regions of behavior can be identified. Above about 600 mK the dissipation saturates, reaching what appears to be a plateau, whilst saturation is also observed for the data below 30 mK. In the intermediate region between 30 and 500 mK fits to the data show that the behavior of both devices is described by a power-law $T^{n}$, with $n=0.50\pm 0.05$ for the 7.95 MHz device and $n=0.49\pm 0.02$ for the 3.87 MHz device. The different sets of measurements of $Q_0^{-1}$ taken at different fields and using different techniques overlay one another in Fig.\ 3a. Estimates for the dissipation in the plateau regions (shown as dashed horizontal lines in Figs.\ 2a and 3a) are obtained by averaging over the five points taken at the highest temperatures for each sample.  Very similar results are obtained for the two devices: $Q^{-1}_{p,s}= 4.1\times10^{-5}$ and $Q^{-1}_{p,l}=3.8\times 10^{-5}$ for the 7.95 MHz and $3.87$ MHz resonators respectively.

The relative shift in the frequency of the resonators as a function of temperature is shown in Figs.\ 2b and 3b. We have plotted the 2 T CW data for the 3.87 MHz resonator as this set provides the best compromise between line-width and signal-to-noise ratio. For both samples, we find that there is an increase in the frequency with temperature over the lower half of the temperature range. The behavior here is well described by a logarithmic dependence, $\delta f/f_0=C_{s(l)} \ln (T/T_0)$ with gradients $C_s=(7.37\pm 0.95)\times10^{-6}$ and $C_l=(8.48\pm 0.30)\times10^{-6}$,  for the 7.95 MHz and $3.87$ MHz resonators respectively. The frequency shift of the 3.87 MHz device stops increasing at about 400 mK and shows a systematic decrease above 600 mK. However, no decrease is seen for the 7.95 MHz sample, though we note that the scatter in the data around 1 K makes it difficult to draw firm conclusions about the underlying behavior at higher temperatures for this device. Discrete jumps in the resonator frequency were observed when the resonator was warmed up (to about 4K) between measurement runs.

The saturation in the dissipation at low temperatures could be due to a variety of factors including a crossover to a regime where a temperature independent mechanism such as clamping losses or Joule heating of the sample dominates.  However, because some saturation is also observed in the frequency shift data, a feature seen clearly in Fig.\ 2, we believe that thermal decoupling  between the sample and the refrigerator is the most likely explanation. Indeed if we assume the logarithmic dependence for the frequency shift in Fig.\ 2b is obeyed below 30 mK and use this as a thermometer for the two points at the lowest temperatures then the corresponding dissipation points recalibrated in this way lie close to the $T^{0.5}$ line.

The strong variation seen in the dissipation at such low temperatures suggests that tunneling two level systems (TLS), which are known to exist in polycrystalline metals,\cite{birge} are the dominant source of the dissipation. The standard model of low temperature acoustic damping in amorphous solids\cite{esqui} assumes tunneling TLS with a broad distribution of energies and relaxation rates. For resonator frequencies, $f$, such that  $hf\ll k_{{\rm B}}T$ (as in our experiments), dissipation is dominated by the relaxation of the TLS (not resonant excitation). A change in the behavior is expected at a temperature $T^*$  where the minimum relaxation time for the TLS matches the period of the resonator. Above $T^{*}$, there is a plateau where the dissipation,\cite{foot,esqui} $Q^{-1}_{p}=\pi C/2$, depends on material properties such as the {spectral} density of TLS, but is independent of temperature and frequency;  below $T^{*}$ the dissipation is expected to have a characteristic power dependence on temperature that depends both on how the TLS relax and the form of the underlying distribution. The dissipation we see in the gold resonators fits this pattern with both devices showing plateau like features above about 600 mK, and a power law dependence $\sim T^{0.5}$ at lower temperatures down to about 30 mK. Interestingly, a recent calculation\cite{seonaz} predicted a $T^{0.5}$ behavior when the relaxation of TLS is due to {\it phonons} in a nanomechanical resonator, {though a plateau region is not predicted in this case}. Furthermore, the dissipation in the plateau regions is very similar for the two devices although it leads to a value $C\sim 2.5\times10^{-5}$ which is about an order of magnitude below the range typical for amorphous solids, {though similar low values are seen\cite{esqui} in some disordered metals as well as in stressed dielectric resonators}.\cite{parpia2}

The standard tunneling model also predicts a frequency shift\cite{esqui} that increases logarithmically with temperature for $T<T^*$ (due to the resonant interaction between the TLS and the acoustic excitation) with the gradient given by $C$. For $T>T^*$ a logarithmic decrease in the frequency shift is expected, with a gradient that depends on the dominant relaxation mechanism. Although we see a logarithmic increase in the frequency shift at low temperatures, the gradients are only about a third of the value of $C$ inferred from the dissipation in the plateau regions. At higher temperatures the frequency shifts for our two resonators show different behaviors and so firm conclusions cannot be drawn.

In summary, we find that the dissipation follows a weak power law dependence $Q_0^{-1}\propto T^{0.5}$ over the range 30 to 500 mK and above 600 mK the dissipation saturates. The relative frequency shift of the resonators shows a logarithmic increase up to about 400 mK. {The observation of features normally associated with tunneling TLS, such as a plateau in the dissipation and a power law behavior at lower temperatures together with the logarithmic change in the frequency shift, suggest that this is the right paradigm in this case. Nevertheless, our results do not fit with current theories. However,} differences between our results and theories based on the behavior of {bulk} amorphous solids are to be expected, not just because the rates at which the TLS relax should be rather different in nanomechanical resonators,\cite{seonaz} but also because the very small volume of the samples\cite{blencowe} together with the effects of tension\cite{parpia2} may lead to a distribution of TLS quite different to that usually assumed for bulk solids.\cite{esqui}

We thank K. Panesar for his help with the ring-down measurements and M. Blencowe for useful discussions. Funding from the EPSRC (UK) under grants EP/E03442X/1 and EP/C540182/1 is acknowledged.


\begin{thebibliography}{99}
\bibitem{ekinci} K. L. Ekinci and M. Roukes, Rev.\ Sci.\ Inst.\ {\bf 76}, 061101 (2005).
\bibitem{naik} A. Naik et al., Nature {\bf 443}, 193 (2006).
\bibitem{lehnert} C. A. Regal, J. D. Teufel, and K. W. Lehnert, Nat.\ Phys.\ {\bf 4}, 555 (2008);
\bibitem{miles} M. Blencowe, Phys.\ Rep.\ {\bf 395}, 159 (2004).
\bibitem{max} M. Schlosshauer, A. P. Hines, and G. J. Milburn, Phys.\ Rev.\ A {\bf 77}, 022111 (2008).
\bibitem{blencowe} L. G. Remus, M. P. Blencowe and Y. Tanaka,  Phys.\ Rev.\ B {\bf 80}, 174103 (2009).
\bibitem{greywall} D. S. Greywall et al., Europhys.\ Lett.\ {\bf 34}, 37 (1996).
\bibitem{mohanty} G. Zolfagharkhani et al., Phys.\ Rev.\ B {\bf 72}, 224101 (2005); P. Mohanty et al.,  Phys.\ Rev.\ B  {\bf 66}, 085416 (2002);  M. Imboden and P. Mohanty, Phys.\ Rev.\ B {\bf 79}, 125424 (2009).
\bibitem{diamond} A. B. Hutchinson et al.,   Appl.\ Phys.\ Lett.\ {\bf 84}, 972 (2004).
\bibitem{ted} R. Lifshitz and M. L. Roukes, Phys.\ Rev.\ B {\bf 61},  5600 (2000).
\bibitem{lc} M. C. Cross and R. Lifshitz, Phys.\ Rev.\ B {\bf 64}, 085324 (2001); D. M. Photiadis and J. A. Judge, Appl.\ Phys.\ Lett. {\bf 85}, 482 (2004).
\bibitem{iwr} I. Wilson-Rae, Phys.\ Rev.\ B {\bf 77}, 245418 (2008).
\bibitem{ahn}{K.-H. Ahn and P. Mohanty, Phys.\ Rev.\ Lett.\ {\bf 90} 085504 (2003).}
\bibitem{seonaz} C. Seo\'{a}nez, F. Guinea and A. H. Castro Neto, Phys.\ Rev.\ B {\bf 77} 125107 (2008).
\bibitem{esqui} P. Esquinazi and R. K\"{o}nig, in  \emph{Tunneling Systems in Amorphous and Crystalline Solids}, ed.\ P. Esquinazi (Springer-Verlag, Berlin, 1998); W. A. Phillips, Rep.\ Prog.\ Phys.\ {\bf 50} 1657 (1987).
\bibitem{cooling} {J. T. Muhonen et al., Appl.\ Phys.\ Lett.\ {\bf 94}, 073101 (2009); P. J. Koppinen and I. J. Maasilta, Phys.\ Rev.\ Lett.\ {\bf 102}, 165502 (2009).}
\bibitem{li}T. F. Li et al., Appl.\ Phys.\ Lett.\ {\bf 92}, 043112 (2008).
\bibitem{venkatesan} A. Venkatesan et al., J.\ Low Temp.\ Phys.\ (proceedings of QFS 2009).
\bibitem{cleland2} A. Cleland and M. L. Roukes, Sensors and Actuators A {\bf 72}, 256 (1999).
\bibitem{roukes} H. W. Ch. Postma et al., Appl.\ Phys.\ Lett.\ {\bf 86} 223105 (2005).
\bibitem{augaas} F. C. Nix and D. MacNair, Phys.\ Rev.\ {\bf 60}, 597 (1941); J. S. Blakemore, J.\ Appl.\ Phys.\ {\bf 53}, R123 (1982).
\bibitem{parpia} S. S. Verbridge et al., Nano Lett.\ {\bf 7}, 1728 (2007).
\bibitem{birge} K. Chun and N. O. Birge, Phys.\ Rev.\ B {\bf 54}, 4629 (1996).
\bibitem{foot} {Such a picture holds for bulk solids where the TLS relaxation is due to either phonons or electrons.}
\bibitem{parpia2} D. R. Southworth et al.,  Phys.\ Rev.\ Lett.\ {\bf 102} 225503 (2009).
\end{thebibliography}
\end{document}